\documentclass[runningheads]{llncs}
\usepackage{graphicx}
\usepackage{amssymb}
\usepackage{amsmath}

\usepackage{times}

\usepackage{caption}
\usepackage{color}
\usepackage{multirow}
\usepackage{xspace}

\newcommand{\graph}{\ensuremath{\mathtt{G}}\xspace}
\newcommand{\txt}{\ensuremath{\mathtt{Text}}\xspace}
\newcommand{\cok}{\ensuremath{\mathtt{coKyw}}\xspace}
\newcommand{\subd}{\ensuremath{\mathtt{SD}}\xspace}
\newcommand{\coc}{\ensuremath{\mathtt{coCnx}}\xspace}
\newcommand{\cls}{\ensuremath{\mathtt{Cls}}\xspace}
\newcommand{\prp}{\ensuremath{\mathtt{Prp}}\xspace}
\newcommand{\frqcls}{\ensuremath{\mathtt{frqCls}}\xspace}
\newcommand{\frqprp}{\ensuremath{\mathtt{frqPrp}}\xspace}
\newcommand{\hm}{\ensuremath{\mathtt{hm}}\xspace}
\newcommand{\cosch}{\ensuremath{\mathtt{coSkm}}\xspace}
\newcommand{\degree}{\ensuremath{\mathtt{d}}\xspace}
\newcommand{\ent}{\ensuremath{\mathtt{Ent}}\xspace}
\newcommand{\codat}{\ensuremath{\mathtt{coDat}}\xspace}

\begin{document}

\title{A Framework for Evaluating Snippet Generation for Dataset Search}

\author{Xiaxia Wang\inst{1} \and
Jinchi Chen\inst{1} \and
Shuxin Li\inst{1} \and
Gong Cheng\inst{1} \and
Jeff Z. Pan\inst{2,3} \and \\
Evgeny Kharlamov\inst{4,5} \and
Yuzhong Qu\inst{1}}

\authorrunning{X. Wang et al.}
\institute{National Key Laboratory for Novel Software Technology, Nanjing University, China\\
\email{
xxwang1997@gmail.com,
\{jcchen,sxli\}@smail.nju.edu.cn,
\{gcheng,yzqu\}@nju.edu.cn
}
\and
 Edinburgh Research Centre, Huawei, UK
\and 
Department of Computing Science, University of Aberdeen, UK\\
\email{
jeff.z.pan@abdn.ac.uk
}
\and
Department of Informatics, University of Oslo, Norway\\
\email{
evgeny.kharlamov@ifi.uio.no
}
\and
Bosch Center for Artificial Intelligence, Robert Bosch GmbH, Germany\\
\email{
evgeny.kharlamov@de.bosch.com
}
}

\maketitle

\begin{abstract}
Reusing existing datasets is of considerable significance to researchers and developers. Dataset search engines help a user find relevant datasets for reuse. They can present a snippet for each retrieved dataset to explain its relevance to the user's data needs. This emerging problem of snippet generation for dataset search has not received much research attention. To provide a basis for future research, we introduce a framework for quantitatively evaluating the quality of a dataset snippet. The proposed metrics assess the extent to which a snippet matches the query intent and covers the main content of the dataset. To establish a baseline, we adapt four state-of-the-art methods from related fields to our problem, and perform an empirical evaluation based on real-world datasets and queries. We also conduct a user study to verify our findings. The results demonstrate the effectiveness of our evaluation framework, and suggest directions for future research.

\end{abstract}

\section{Introduction}

We are witnessing the rapid growth of open data on the Web, notably RDF, Linked Data and Knowledge Graphs~\cite{Pan2016}. Today, to develop a Web application, reusing existing datasets not only brings about productivity improvements and cost reductions, but also makes interoperability with other applications more achievable. However, there is a lack of tool support for conveniently finding datasets that match a developer's data needs. To address it, recent research efforts yielded \emph{dataset search engines} like LODAtlas~\cite{lodatlas} and Google Dataset Search~\cite{DBLP:conf/www/BrickleyBN19}. They retrieve a list of datasets that are relevant to a keyword query by matching the query with the description in the metadata of each dataset.

These systems have made a promising start.
%In addition to retrieval,
Furthermore, a helpful dataset search engine
%in addition to retrieving datasets relevant to a user's data needs
should also explain why a retrieved dataset is relevant. A concise piece of information presented for each dataset in a search results page is broadly referred to as a \emph{dataset summary}. It may help the user quickly identify a relevant dataset. Summaries presented in current dataset search engines, however,  are mainly composed of some \emph{metadata} about a dataset, such as provenance and license. Their utility in relevance judgment is limited,
%because the information they span is orthogonal to the query and the content of the dataset.
with users  having to analyze each dataset in the search results to assess its relevance, which would be a time-consuming process.

To overcome the shortcoming of metadata, we study an emerging type of dataset summary called \emph{dataset snippet}. For an RDF dataset retrieved by a keyword query, a dataset snippet is a size-constrained subset of RDF triples extracted from the dataset, being intended to exemplify the content of the dataset and to explain its relevance to the query. It differs from a \emph{dataset profile} which represents a set of features describing attributes of the dataset~\cite{DBLP:journals/semweb/EllefiBBDDST18}. It is also complementary to an \emph{abstractive summary} which aggregates data into patterns and provides a high-level overview~\cite{DBLP:journals/pvldb/CebiricGM15,hieds,DBLP:journals/tkde/SongWLDS18,rdfdigest,DBLP:journals/semweb/ZneikaVK19}. It is conceptually more similar to a snippet extracted from a webpage and presented in traditional Web search. However, little research attention has focused on this perspective.

As a preliminary effort along this way, we work towards establishing a framework for evaluating snippets generated for dataset search. That would provide a basis for future research, in terms of providing quantitative evaluation metrics and advising algorithm design. Existing evaluation metrics used in related fields such as snippet generation for ontologies~\cite{ontosum1} and documents~\cite{dssurvey} are mainly based on a human-created ground truth. However, an RDF dataset may contain millions of RDF triples, 
e.g., when it wrapped from a large database~\cite{DBLP:journals/ws/KharlamovHSBJXS17,DBLP:journals/internet/HorrocksGKW16,DBLP:conf/semweb/Jimenez-RuizKZH15,DBLP:journals/semweb/PinkelBJKMNBSST18},
or streaming data~\cite{DBLP:journals/ws/KharlamovMMNORS17,KHARLAMOV201930}, or comes from a manufacturing environment~\cite{DBLP:conf/esws/RingsquandlK0HL18,DBLP:conf/semweb/KharlamovGJLMRN16a,KHARLAMOV201911} being much larger than an ontology schema or a document. It would be difficult, if not impossible, to manually identify the optimum snippet as the ground truth. Therefore, new evaluation metrics are needed.

To demonstrate the use of our evaluation framework, considering the lack of dedicated solutions to dataset snippets, we explore research efforts in related fields and adapt their methods to our problem. Using our framework, we analyze these methods and empirically evaluate them based on real-world datasets. We also carry out a user study to verify our findings and solicit comments to motivate future research.

To summarize, our contributions in this paper include
\begin{itemize}
    \item a framework for evaluating snippets in dataset search, consisting of four metrics regarding how well a snippet covers a query and a dataset,
    \item an adaptation of four state-of-the-art methods selected from related fields to generate snippets for dataset search, as a baseline for future research, and
    \item an evaluation of the adapted methods using the proposed evaluation framework based on real-world datasets and queries, as well as a user study.
\end{itemize}

The remainder of the paper is organized as follows. Section~\ref{sec:rw} reviews related research. Section~\ref{sec:fw} describes our evaluation framework. Section~\ref{sec:eva} reports evaluation results. Section~\ref{sec:us} presents a user study. Section~\ref{sec:cfd} concludes the paper.
\section{Related Work}\label{sec:rw}

Very little research attention has been given to the problem of snippet generation for dataset search. Therefore, in this section, we also review research efforts in related fields that can be adapted to the problem we study.

\subsection{Snippets for RDF Datasets}\label{sec:rw-r}

In an early work~\cite{otm}, a snippet for an RDF document is generated to show how the document is relevant to a keyword query. Preference is given to RDF triples that describe central entities or contain query keywords. The proposed algorithm relies on manually defined ranking of predicates. In~\cite{DBLP:conf/aaai/DolbyFKKSSM07,DBLP:conf/semweb/RietveldHSG14}, an RDF dataset is compressed by keeping only a sample of triples in order to improve the performance of query processing while still serve query results as complete as possible. To this end, \cite{DBLP:conf/semweb/RietveldHSG14}~samples triples that are central in the RDF graph and hence are likely to appear in the answers of typical SPARQL queries. By contrast, \cite{DBLP:conf/aaai/DolbyFKKSSM07}~iteratively expands the sample as needed to make it more precise. Completeness preserving summaries~\cite{FMSP2012} help optimise distributed reasoning and querying.

In a recent work~\cite{wsdm}, an \emph{illustrative snippet} is generated to exemplify the content of an RDF dataset. Snippet generation is formulated as a combinatorial optimization problem, aiming to find an optimum connected RDF subgraph such that it contains instantiation of the most frequently used classes and properties in the dataset and contains entities having the highest PageRank scores. An approximation algorithm is presented to solve this NP-hard problem. This kind of snippet can be used in dataset search, although it is not query-biased.

\subsection{Snippets for Ontology Schemas}\label{sec:rw-o}

An \emph{ontology snippet} distills the most important information from an ontology schema and forms an abridged version~\cite{DBLP:conf/www/ZhangCQ07,DBLP:journals/jcst/ZhangCGQ09}. Existing methods often represent an ontology schema as a graph, and apply some centrality-based measures to identify the most important terms or axioms as an ontology snippet~\cite{aike,ijsc}. It is possible to adapt these methods to generate snippets for an RDF dataset because it can be viewed as an RDF graph to process.

We give particular attention to methods that are capable of generating \emph{query-biased snippets for ontology search}~\cite{dwrank,ipm,aswc,DBLP:conf/www/ChengGQ11,biprank}. An ontology schema is often represented as a graph where nodes represent terms and edges represent axioms associating terms~\cite{DBLP:conf/aswc/ZhangLQ06,ipm}.
%~\cite{DBLP:journals/ws/ArenasGKMZ16,DBLP:journals/semweb/SoyluKZJGSHSBLH18,DBLP:journals/uais/SoyluGJKZH17,DBLP:conf/cikm/KharlamovGSGKH17,DBLP:conf/semweb/SherkhonovGKK17}
In a state-of-the-art approach~\cite{ipm}, such a graph is decomposed into a set of maximal radius-bounded connected subgraphs, which in turn are reduced to tree-structured sub-snippets. A greedy algorithm is performed to select and merge an optimum set of sub-snippets, in terms of compactness and query relevance.

\subsection{Keyword Search on Graphs}\label{sec:rw-k}

Keyword search on a graph is to find an optimum connected subgraph that contains all the keywords in a query~\cite{kwsurvey,DBLP:conf/bigdataconf/ChengK17}.
An optimum subgraph has the smallest total edge weight~\cite{dpbf,star,DBLP:conf/sigmod/LiQYM16},
%banks
or a variant of this property~\cite{DBLP:journals/tkde/LeLKD14}. As each keyword can match a set of nodes in a graph, the problem is formulated as a \emph{group Steiner tree (GST) problem}. This kind of subgraph can be used as a query-biased snippet for an RDF dataset viewed as an RDF graph. However, the problem is NP-hard and is difficult to solve. Many algorithms perform not well on large graphs~\cite{kwexp}.

A state-of-the-art algorithm for the GST problem is PrunedDP++~\cite{DBLP:conf/sigmod/LiQYM16}. The algorithm progressively refines feasible solutions based on dynamic programming with an A*-search strategy. In dynamic programming, optimal-tree decomposition and conditional tree merging techniques are proposed to prune unpromising states. For A*-search, several lower-bounding techniques are used.

\subsection{Snippets for Documents}\label{sec:rw-d}

%Document summarization has been studied for decades. A document summary can be generated through extractive or abstractive methods.
A \emph{document snippet} consists of salient sentences selected from the original document~\cite{dssurvey}.
%whereas abstractive methods generate new sentences and are much more complex.
To adapt such a method to our problem, we could replace the three elements of an RDF triple with their textual forms. The triple becomes a pseudo sentence, and an RDF dataset is transformed into a set of sentences to process.

Among existing solutions, \emph{unsupervised query-biased methods}~\cite{qbds2}
% qbds1
are closer to our problem setting because, at this stage, training data for dataset search is not available. The CES method~\cite{DBLP:conf/sigir/FeigenblatRBK17} is among the state-of-the-art in this line of work. It formulates sentence selection as an optimization problem and solves it using the cross-entropy method. Preference is given to diversified long sentences that are relevant to a query.
\section{Evaluation Framework}\label{sec:fw}

In this section, we firstly define some terms used in the paper, and then propose a framework for evaluating snippets generated for dataset search. Our framework, consisting of four metrics characterizing different aspects of a dataset snippet, will be used in later sections to evaluate selected methods reviewed in Section~\ref{sec:rw}.

\subsection{Preliminaries}

Datasets vary in their formats.
%e.g., CSV files, XML documents.
Search queries have various types.
%, e.g., natural language queries, SPARQL queries.
This paper is focused on keyword queries over RDF datasets because this combination is common. We will consider other data formats and query types in future work.
\begin{definition}[RDF Dataset]
  An RDF dataset, or a dataset for short, is a set of $n$~RDF triples denoted by $T=\{t_1,\ldots,t_n\}$. Each $t_i \in T$ is a subject-predicate-object triple denoted by $\langle t_i^\text{s},t_i^\text{p},t_i^\text{o} \rangle$.
\end{definition}
\noindent In RDF, $t_i^\text{s}$, $t_i^\text{p}$, and~$t_i^\text{o}$ can be IRIs, blank nodes, or literals, which are collectively known as RDF terms. An ``RDF term'' and the ``resource'' it denotes are used interchangeably in the paper.
\begin{definition}[Keyword Query]
  A keyword query, or a query for short, is a set of $m$~keywords denoted by $Q=\{q_1,\ldots,q_m\}$.
\end{definition}

A snippet of a dataset is a size-constrained subset of triples extracted from the dataset. The extraction should consider the query.
\begin{definition}[Dataset Snippet]
  Given a positive integer~$k$, a snippet of a dataset~$T$ is denoted by~$S$ subject to $S \subseteq T$ and $|S| \leq k$.
\end{definition}

\begin{figure}[!t]
    \centering
    \includegraphics[width=\textwidth]{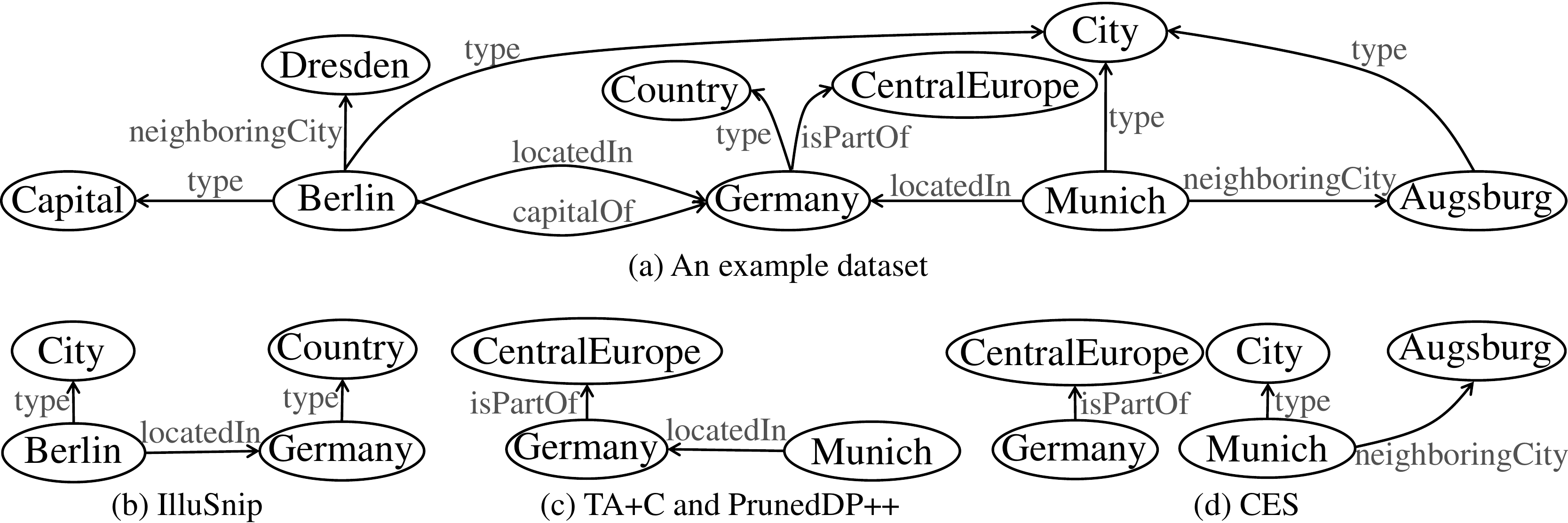}
    \caption{(a)~An example dataset and (b)(c)(d)~three of its snippets generated by different methods w.r.t. the query \emph{munich europe}.}
    \label{fig:toydataset}
\end{figure}

An RDF dataset~$T$ can be viewed as an RDF graph denoted by $\graph(T)$. Each triple $\langle t^\text{s},t^\text{p},t^\text{o} \rangle \in T$ is represented as a directed edge labeled with~$t^\text{p}$ from node~$t^\text{s}$ to node~$t^\text{o}$ in $\graph(T)$. Analogously, a snippet~$S$ is a subgraph denoted by $\graph(S)$. In Fig.~\ref{fig:toydataset} we illustrate three snippets for an example dataset w.r.t. a query.

\subsection{Evaluation Metrics}

To assess the quality of a snippet w.r.t. a query, we propose four quantitative metrics: $\cok$, $\coc$, $\cosch$, and $\codat$. Recall that a snippet is generated to exemplify the content of a dataset and to explain its relevance to the query. So a good snippet should, on the one hand, match the query intent ($\cok$, $\coc$) and, on the other hand, cover the main content of the dataset ($\cosch$, $\codat$).
%Two of them measure the snippet's coverage of the query, and the other two measure the snippet's coverage of the dataset.
Our metrics are open source\footnote{\url{http://ws.nju.edu.cn/datasetsearch/evaluation-iswc2019/metrics.zip}}.

\subsubsection{Coverage of Query Keywords ($\cok$).}

Keywords in a query express a user's data needs. A good snippet should cover as many keywords as possible, to show how a dataset is plainly relevant to the query.

Specifically, let $\txt(r)$ be a set of textual forms of a resource~$r$. For~$r$ denoted by an IRI, $\txt(r)$ include
\begin{itemize}
    \item the lexical forms of $r$'s \emph{human-readable names} (if any), i.e., literal values of $r$'s \texttt{rdfs:label} property, and
    \item $r$'s \emph{local name}, i.e., the fragment component of $r$'s IRI (if its exists) or the last segment of the path component of the IRI.
\end{itemize}
\noindent For~$r$ denoted by a blank node, $\txt(r)$ only include the lexical forms of $r$'s human-readable names (if any). For~$r$ denoted by a literal, $\txt(r)$ only include the lexical form of the literal.

A resource~$r$ \emph{covers} a keyword~$q$ if any textual form in~$\txt(r)$ contains a \emph{match} for~$q$. Our implementation considers keyword matching, which can be extended to semantic matching in future work. A triple~$t$ \emph{covers} a keyword~$q$, denoted by $t \prec q$, if $r$~covers~$q$ for any $r \in \{t^\text{s},t^\text{p},t^\text{o}\}$. For a snippet~$S$, its coverage of keywords in a query~$Q$ is the proportion of covered keywords:
\begin{equation}
    \cok(S) = \frac{1}{|Q|} \cdot |\{q \in Q : \exists t \in S, ~t \prec q\}| \,.
\end{equation}

For example, Fig.~\ref{fig:toydataset}(c) and~(d) cover all the query keywords, so $\cok=1$. None of the keywords are covered by Fig.~\ref{fig:toydataset}(b), so $\cok=0$.

\subsubsection{Coverage of Connections between Query Keywords ($\coc$).}

Keywords in a query are not independent but often refer to a set of related concepts which collectively represent a query intent. To show how a dataset is relevant to the query and its underlying intent, a good snippet should cover not only query keywords but also their connections captured by the dataset.

Specifically, for a snippet~$S$, consider its RDF graph $\graph(S)$. Query keywords can be covered by nodes or edges of $\graph(S)$. For convenience, we obtain a \emph{subdivision} of $\graph(S)$, by subdividing every edge labeled with~$t^\text{p}$ from node~$t^\text{s}$ to node~$t^\text{o}$ into two unlabeled undirected edges: one between~$t^\text{s}$ and~$t^\text{p}$, and the other between~$t^\text{p}$ and~$t^\text{o}$. The resulting graph is denoted by $\subd(\graph(S))$. A snippet~$S$ \emph{covers} the connection between two keywords $q_i,q_j \in Q$, denoted by $S \prec (q_i,q_j)$, if there is a path in $\subd(\graph(S))$ that connects two nodes: one covering~$q_i$ and the other covering~$q_j$. For~$S$, its coverage of connections between keywords in~$Q$ is the proportion of covered connections between unordered pairs of keywords:
\begin{equation}\label{eq:coc}
    \coc(S) =
    \begin{cases}
      \frac{1}{\binom{|Q|}{2}} \cdot |\{\{q_i,q_j\} \subseteq Q : q_i \neq q_j \text{ and } S \prec (q_i,q_j)\}| & \text{if } |Q|>1 \,,\\
      \cok(S) & \text{if } |Q|=1 \,.
    \end{cases}
\end{equation}
\noindent When there is only one keyword, $\coc$ is meaningless and we set it to $\cok$.

%Further constraints can be placed on the above path to only accept certain kinds of connections. In our implementation, \emph{entity-level connections} are our focus. Entities are non-literal resources at the data level. Therefore, two kinds of paths are ignored when identifying connections: those passing through literals and those passing through the \texttt{rdf:type} property, which represent literal-level and class-level connections, respectively.

For example, Fig.~\ref{fig:toydataset}(c) covers the connection between the two query keywords, so $\coc=1$. By contrast, although Fig.~\ref{fig:toydataset}(d) covers all the keywords, it fails to cover their connections, so $\coc=0$.

\subsubsection{Coverage of Data Schema ($\cosch$).}

Snippets are expected to not only interpret query relevance but also offer a representative preview of a dataset. In particular, the RDF schema of a dataset is important to users. A good snippet should cover as many classes and properties used in the dataset as possible, to exemplify which types of things and facts a user can obtain from the dataset.

Specifically, a snippet~$S$ \emph{covers} a class or a property if $S$~contains its instantiation. Let $\cls(S)$ and $\prp(S)$ be the set of classes and the set of properties instantiated in~$S$, respectively:
\begin{equation}\label{eq:clsprp}
\begin{split}
  \cls(S) & = \{c : \exists t \in S, ~t^\text{p}=\texttt{rdf:type} \text{ and } t^\text{o}=c\} \,,\\
  \prp(S) & = \{p : \exists t \in S, ~t^\text{p}=p\} \,.
\end{split}
\end{equation}

Classes and properties that are used more often in a dataset are more representative. The relative frequency of a class~$c$ observed in a dataset~$T$ is
\begin{equation}
  \frqcls(c) = \frac{|\{t \in T : ~t^\text{p}=\texttt{rdf:type} \text{ and } t^\text{o}=c\}|}{|\{t \in T : ~t^\text{p}=\texttt{rdf:type}\}|} \,.
\end{equation}
\noindent Analogously, the relative frequency of a property~$p$ observed in~$T$ is
\begin{equation}
  \frqprp(p) = \frac{|\{t \in T : ~t^\text{p}=p\}|}{|T|} \,.
\end{equation}
\noindent For a snippet~$S$, its coverage of the schema of~$T$ is related to: (a)~the total relative frequency of the covered classes, and (b)~the total relative frequency of the covered properties. We calculate the harmonic mean~($\hm$) of the two:
\begin{equation}
\begin{split}
  \cosch(S) & = \hm(\sum_{c \in \cls(S)}{\frqcls(c)}, ~\sum_{p \in \prp(S)}{\frqprp(p)}) \,,\\
  \hm(x,y) & = \frac{2xy}{x+y} \,.
\end{split}
\end{equation}

For example, Fig.~\ref{fig:toydataset}(b) covers a frequent class (\texttt{City}) and a frequent property (\texttt{locatedIn}) in the dataset, so its $\cosch$ score is higher than that of Fig.~\ref{fig:toydataset}(c) which covers only properties but not classes.
%$0.709$, $0$, and~$0.658$, respectively.

\subsubsection{Coverage of Data ($\codat$).}

Classes and properties of high relative frequency are central elements in the schema used in a dataset. Complementary to them, a good snippet should also cover central elements at the data level (i.e., central entities), to show the key content of the dataset.

Specifically, let $\degree^+(r)$ and $\degree^-(r)$ be the out-degree and in-degree of a resource~$r$ in an RDF graph $\graph(T)$, respectively:
\begin{equation}
\begin{split}
  \degree^+(r) = |\{t \in T : t^\text{s}=r\}| \,,\\
  \degree^-(r) = |\{t \in T : t^\text{o}=r\}| \,.
\end{split}
\end{equation}
\noindent Out-degree characterizes the richness of the description of a resource, and in-degree characterizes popularity. They suggest the centrality of a resource from different aspects. For a snippet~$S$, its coverage of a dataset~$T$ at the data level is related to: (a)~the mean normalized out-degree of the constituent entities, and (b)~the mean normalized in-degree of the constituent entities. We calculate the harmonic mean of the two:
\begin{equation}
\begin{split}
  \codat(S) & = \hm(\frac{1}{|\ent(S)|} \cdot \sum_{e \in \ent(S)}{\frac{\log (\degree^+(e)+1)}{\max_{e' \in \ent(T)}{\log (\degree^+(e')+1)}}}, \\
  & \quad\quad\quad \frac{1}{|\ent(S)|} \cdot \sum_{e \in \ent(S)}{\frac{\log (\degree^-(e)+1)}{\max_{e' \in \ent(T)}{\log (\degree^-(e')+1)}}}) \,,\\
  \ent(X) & = \{r : \exists t \in X, ~r \in \{t^\text{s},t^\text{o}\}, ~r \notin \cls(T), ~\text{and $r$~is not a literal.}\} \,,
\end{split}
\end{equation}
\noindent where $\cls(T)$ is the set of all classes instantiated in~$T$ defined in Eq.~(\ref{eq:clsprp}), $\ent(S)$ is the set of all entities (i.e., non-literal resources at the data level) that appear in~$S$, and $\ent(T)$ is the set of all entities that appear in~$T$. Degree is normalized by the maximum value observed in the dataset. Considering that degree usually follows a highly skewed power-law distribution in practice, normalization is performed on a logarithmic scale.

For example, Fig.~\ref{fig:toydataset}(b) is focused on \texttt{Germany}, which is a central entity in the dataset, so its $\codat$ score is higher than that of Fig.~\ref{fig:toydataset}(c) and~(d) which contain more of subordinate entities.
% $0.617$, $0.480$, and~$0.470$, respectively.
\section{Evaluation}\label{sec:eva}

In Section~\ref{sec:rw}, each subsection reviews methods in a related research field that can be adapted to generate snippets for dataset search. The second paragraph of each subsection identifies a state-of-the-art method from each field that is suitable for our context: \cite{wsdm}, \cite{ipm}, \cite{DBLP:conf/sigmod/LiQYM16}, and~\cite{DBLP:conf/sigir/FeigenblatRBK17}. In this section, we evaluate these methods using the evaluation framework proposed in Section~\ref{sec:fw}. We first analyze whether and how the components of these methods are aligned with each evaluation metric. Then we perform an extensive empirical evaluation based on real-world datasets.

\begin{table}[!t]
\centering
\caption{Overview of selected methods and their alignment with evaluation metrics.}\label{tab:teva}
\begin{tabular}{|ll|c|c|c|c|}
%\begin{tabular}{|p{6.4cm}|p{1.4cm}|p{1.35cm}|p{1.3cm}|p{1.3cm}|}
\hline
%\multirow{2}{*}{} & \multicolumn{2}{c|}{Coverage of query} & \multicolumn{2}{c|}{Coverage of dataset} \\
%\cline{2-5}
& & $\cok$ & $\coc$ & $\cosch$ & $\codat$ \\
%& Keywords ($\cok$) & Conn. between keywords ($\coc$) & Schema level ($\cosch$) & Data level ($\codat$) \\
\hline
IlluSnip~\cite{wsdm} & (illustrative dataset snippet) & & & $\checkmark$ & $\checkmark$ \\
TA$+$C~\cite{ipm} & (query-biased ontology snippet) & $\checkmark$ & $\checkmark$ & & \\
PrunedDP$++$~\cite{DBLP:conf/sigmod/LiQYM16} & (GST for keyword search) & $\checkmark$ & $\checkmark$ & & \\
CES~\cite{DBLP:conf/sigir/FeigenblatRBK17} & (query-biased document snippet) & $\checkmark$ & & $\checkmark$ & $\checkmark$ \\
\hline
\end{tabular}
\end{table}

\subsection{Analysis of Selected Methods}\label{sec:teva}

Table~\ref{tab:teva} presents an overview of the selected methods and whether they have components that are conceptually similar to each evaluation metric. All the methods have been detailed in Section~\ref{sec:rw}. In the following we focus on how their components are aligned with each evaluation metric.

\subsubsection{Illustrative Dataset Snippet.}

Dataset snippets generated by existing methods reviewed in Section~\ref{sec:rw-r} can be used in dataset search without adaptation. The method we choose, IlluSnip~\cite{wsdm}, generates an illustrative snippet for an RDF dataset by extracting a connected subgraph to exemplify the content of the dataset. This intended use is very close to our problem.
% The subgraph contains frequently used classes, properties, and central entities in the dataset.

IlluSnip explicitly considers a snippet's coverage of a dataset. Giving priority to the most frequent classes and properties, a snippet is likely to show a high coverage of data schema~($\cosch$). Besides, IlluSnip computes the centrality of an entity by PageRank, which positively correlates with in-degree. Therefore, a snippet containing such central entities may also have a reasonably high coverage of data~($\codat$), which is jointly measured by in-degree and out-degree.

However, IlluSnip is not query biased. A snippet it generates may not contain any keyword in a query, and hence its coverage of query keywords~($\cok$) and the connections thereof~($\coc$) can be very low.

For example, Fig.~\ref{fig:toydataset}(b) illustrates a snippet generated by IlluSnip.

\subsubsection{Query-biased Ontology Snippet.}

Query-biased snippets for ontology search reviewed in Section~\ref{sec:rw-o} are useful for deciding the relevance of an ontology schema to a query. It is similar to our intent to support judging the relevance of a dataset. The method we choose, TA$+$C~\cite{ipm}, extracts a query-biased subgraph from the RDF graph representation of an ontology schema. This method can be directly used to generate snippets for RDF datasets without adaptation.

TA$+$C explicitly considers a snippet's coverage of a query. It greedily adds query-biased sub-snippets into a snippet, giving preference to those containing more query keywords. A sub-snippet is a radius-bounded connected subgraph. Therefore, the resulting snippet has the potential to establish a high coverage of query keywords~($\cok$) and their connections~($\coc$), especially when keywords are closely located in the dataset.

On the other hand, coverage of dataset ($\cosch$ and $\codat$) is not of concern to this query-centered method.

For example, Fig.~\ref{fig:toydataset}(c) illustrates a snippet generated by TA$+$C.

\subsubsection{GST for Keyword Search.}

Methods for keyword search on graphs reviewed in Section~\ref{sec:rw-k} find a GST, which is a connected subgraph where nodes contain all the query keywords. These methods can be straightforwardly applied to generate snippets for RDF datasets by computing a GST. The method we choose, PrunedDP$++$~\cite{DBLP:conf/sigmod/LiQYM16}, is one of the most efficient algorithms for the GST problem.

PrunedDP$++$ has two possible outputs. Either it finds a GST that covers all the query keywords~($\cok$) and connections between all pairs of them~($\coc$), or such a GST does not exist. In the latter case, PrunedDP$++$ returns empty results. So it is conceptually similar to TA$+$C but appears more ``aggressive''.

Coverage of dataset ($\cosch$ and $\codat$) is not the focus of PrunedDP$++$. Nevertheless, these factors may be partially addressed by properly defining edge weights. Weighting is orthogonal to the design of PrunedDP$++$.

For example, Fig.~\ref{fig:toydataset}(c) illustrates a snippet generated by PrunedDP$++$.

\subsubsection{Query-biased Document Snippet.}

Query-biased methods for generating document snippets reviewed in Section~\ref{sec:rw-d} can be adapted to generate snippets for RDF datasets, by replacing resources in a triple with their textual forms (e.g., labels of IRI-identified resources, lexical forms of literals) to obtain a pseudo sentence. The method we choose, CES~\cite{DBLP:conf/sigir/FeigenblatRBK17}, generates a query-biased snippet by selecting a subset of sentences (i.e., triples in our context). This unsupervised method fits current dataset search, for which training data is in shortage.

CES tends to select diversified triples that are relevant to a query, so it is likely to achieve a high coverage of query keywords~($\cok$). CES also computes the cosine similarity between the term frequency---inverse document frequency (TF-IDF) vectors of the document (i.e., RDF dataset) and a snippet. This feature measures to what extent the snippet covers the main content of the dataset. It increases the possibility of including frequent classes, properties, and entities, and hence may improve a snippet's coverage of dataset ($\cosch$~and~$\codat$).

As a side effect of diversification, triples in a snippet are usually disparate. Connections between query keywords~($\coc$) can hardly be observed.

For example, Fig.~\ref{fig:toydataset}(d) illustrates a snippet generated by CES.
\subsection{Empirical Evaluation}\label{sec:eeva}

We used the proposed framework to evaluate the above selected methods.
%based on real-world RDF datasets
All the experiments were performed on an Intel Xeon E7-4820 (2.00 GHz) with 80GB memory for the JVM. Our implementation of these methods is open source\footnote{\url{http://ws.nju.edu.cn/datasetsearch/evaluation-iswc2019/baselines.zip}}.

\subsubsection{Datasets and Queries.}

We retrieved the metadata of 11,462~datasets from DataHub\footnote{\url{https://old.datahub.io/}} using CKAN's API. Among 1,262~RDF datasets that provided Turtle, RDF/XML, or N-Triples dump files, we downloaded and parsed 311~datasets using Apache Jena. The others were excluded due to download or parse errors.

We used two kinds of queries: real queries and artificial queries.

\emph{Real Queries.}
We used crowdsourced natural language queries\footnote{\url{https://github.com/chabrowa/data-requests-query-dataset}} that were originally submitted to \url{data.gov.uk} for datasets~\cite{DBLP:journals/ws/KacprzakKIBTS19}. They were transformed into keyword queries by removing stop words using Apache Lucene.

\emph{Artificial Queries.}
To have more queries, we leveraged the DMOZ open directory\footnote{\url{http://dmoz-odp.org/}} to imitate possible data needs. For each $i = 1 \ldots 4$, we constructed a group of queries denoted by DMOZ-$i$. A query in DMOZ-$i$ consisted of the names of $i$~random sub-categories of a random top-level category in DMOZ. Such closely related concepts had a reasonable chance to be fully covered by some dataset.

To conclude, we had five groups of queries: data.gov.uk, DMOZ-1, DMOZ-2, DMOZ-3, and DMOZ-4. For each group, we randomly retained $100$~queries such that each query could be paired with a dataset that covered all the query keywords. These $500$~query-dataset pairs were used in our evaluation. We required a dataset to cover all the query keywords in order to make sense of the experiment results.
%For a dataset, its relevance to the query establishes an 'upper bound’ on the query coverage metrics (coKyw, coCnx) of its snippet.
Otherwise, a low score of $\cok$ would be ambiguous: reflecting the poor quality of the snippet, and/or the irrelevance of the dataset.
% For each query-dataset pair, each method generated a snippet for the dataset, possibly taking the query into consideration (depending on the method).

\subsubsection{Configuration of Methods.}

We detail their configuration in the following.

\emph{Size of Snippet.}
Following~\cite{wsdm}, we configured IlluSnip and CES to generate a snippet containing at most 20~RDF triples (i.e., $k=20$). For TA$+$C, it would be inappropriate to bound the number of triples because the snippets it generated could contain isolated nodes. So we bounded it to output a snippet whose graph representation contained at most 20~nodes. For PrunedDP$++$, the size of its output was automatically determined but not configurable.

\emph{Weights and Parameters.}
For TA$+$C~\cite{ipm}, edge weights were defined as in the original paper.
% The weight of edge was set to reciprocal of the number of edges between two nodes.
For PrunedDP$++$~\cite{DBLP:conf/sigmod/LiQYM16}, its authors did not specify how to weight edges. Our weighting followed~\cite{dpbf}---the predecessor of PrunedDP$++$.
% OLD: The weight of edge was set to $log(n+1)$, where n meant the number of edges between two nodes.
For CES~\cite{DBLP:conf/sigir/FeigenblatRBK17}, it had several parameters. Most of them were set to the values used in the original paper. However, due to the large size of RDF dataset, the sampling step in CES was performed 1,000~times (instead of 10,000~times in~\cite{DBLP:conf/sigir/FeigenblatRBK17}) in consideration of memory use.
%TF-IDF was calculated on Wikipedia.

\emph{Preprocessing.}
We built inverted indexes for efficient keyword mapping in TA$+$C, PrunedDP$++$, and CES. For TA$+$C, following its original implementation~\cite{ipm}, we precomputed and materialized all the maximal 1-radius subgraphs.
% \textbf{TA$+$C} In the pre-processing phase, maximal 1-radius subgraphs were computed and stored in the database. Also, two inverted indices were built: one mapped keywords to subgraphs and the other mapped keywords and subgraphs to nodes.
% \textbf{PrunedDP$++$} In the pre-processing phase, keywords inverted index was built.
% \textbf{CES} In the pre-processing phase, index which mapped keywords to triples was built. When the result converged, triples with highest probability would be chosen.

\emph{Timeout.}
After preprocessing, we set a timeout of one hour for each method to generate a snippet. The generating process would be terminated when reaching timeout. In that case, the runtime would be defined to be one hour. For IlluSnip and CES which iteratively found better snippets, the best snippet at timeout would be returned. For TA$+$C and PrunedDP$++$, timeout indicated failure.

\begin{table}[!t]
\centering
\caption{Statistics about query-dataset (Q-D) pairs.}\label{tab:prnmbr}
\resizebox{\textwidth}{!}{
\begin{tabular}{|l|c|c|c|c|c|c|c|c|c|}
  \hline
  \multirow{2}{*}{} & \#Q-D & \multicolumn{2}{|c|}{\#keywords in Q} & \multicolumn{2}{|c|}{\#triples in D} & \multicolumn{2}{|c|}{\#classes in D} & \multicolumn{2}{|c|}{\#properties in D} \\
  \cline{3-10}
  & pairs & mean & max & mean & max & mean & max & mean & max \\
  \hline
  data.gov.uk & 42 & 2.88  & 8 & 116,822 & 2,203,699 & 13 & 129 & 47 & 357 \\
  DMOZ-1 & 88 & 1.25  & 3 & 137,257 & 2,203,699 & 30 & 2,030 & 66 & 3,982 \\
  DMOZ-2 & 84 & 2.33  & 5 & 151,104 & 2,203,699 & 10 & 129 & 34 & 357 \\
  DMOZ-3 & 87 & 3.66  & 6 & 164,714 & 2,203,699 & 13 & 153 & 43 & 357 \\
  DMOZ-4 & 86 & 5.02  & 8 & 219,844 & 2,203,699 & 13 & 129 & 46 & 357 \\
  \hline
\end{tabular}
}
\end{table}
%average class number, average object-type property number, average data-type property number: 
%data.gov.uk: 13, 25, 22
%DMOZ-1: 30, 11, 55
%DMOZ-2: 10, 18, 16
%DMOZ-3: 13, 23, 20
%DMOZ-4: 13, 26, 21

\subsubsection{Evaluation Results.}

Out of the 500~query-dataset pairs, 113~pairs were not included in our results for one of the following reasons.
\begin{itemize}
    \item PrunedDP$++$ did not find any GST to connect all the query keywords, and hence generated an empty snippet.
    \item TA$+$C and PrunedDP$++$ were forced to terminate due to timeout.
    \item TA$+$C did not complete preprocessing after twelve hours.
\end{itemize}
\noindent We reported evaluation results on the remaining 387~pairs where every method successfully generated a non-empty snippet before timeout. Table~\ref{tab:prnmbr} characterizes these queries and datasets. They are available online\footnote{\url{http://ws.nju.edu.cn/datasetsearch/evaluation-iswc2019/query-dataset-pairs.zip}}.
% Among them, the method IlluSnip reaches timeout on $29$ datasets, TA$+$C reaches timeout or returns empty results on $7$ pairs, PrunedDP$++$ returns empty results on $73$ pairs, and CES reaches timeout on $72$ pairs.

Note that IlluSnip and CES were configured to generate a snippet containing at most 20~triples, and they selected~19.68 and 19.89~triples on average, respectively. By comparison, for PrunedDP$++$ the size of its output was automatically determined, and the mean number of triples in the experiment was only~4.60. This may affect the evaluation results. Besides, TA$+$C and PrunedDP$++$ sometimes produced isolated nodes instead of triples. The query keywords covered by these nodes were considered in the computation of $\cok$ and $\coc$.
% IlluSnip, TA$+$C, PrunedDP$++$, and CES were $19.68$, $0.77$, $4.60$, and $19.89$, respectively
% TA+C mean node number: 11.90

\begin{figure}[!t]
    \centering
    \begin{minipage}{.5\textwidth}
        \centering
        \captionsetup{type=table}
        \begin{tabular}{|l|c|c|c|c|}
            \hline
            & $\cok$ & $\coc$ & $\cosch$ & $\codat$ \\
            \hline
            IlluSnip & 0.1000 & 0.0540 & 0.6820 & 0.3850 \\
            TA$+$C & 0.9590 & 0.4703 & 0.0425 & 0.0915 \\
            PrunedDP$++$ & 1 & 1 & 0.0898 & 0.2133 \\
            CES & 0.9006 & 0.3926 & 0.3668 & 0.2684 \\
            \hline
        \end{tabular}
        \caption{Average scores of evaluation metrics on all the query-dataset pairs.}
        \label{tab:avgsc}
    \end{minipage}%
    \hfill
    \begin{minipage}{0.4\textwidth}
        \centering
        \includegraphics[width=\textwidth]{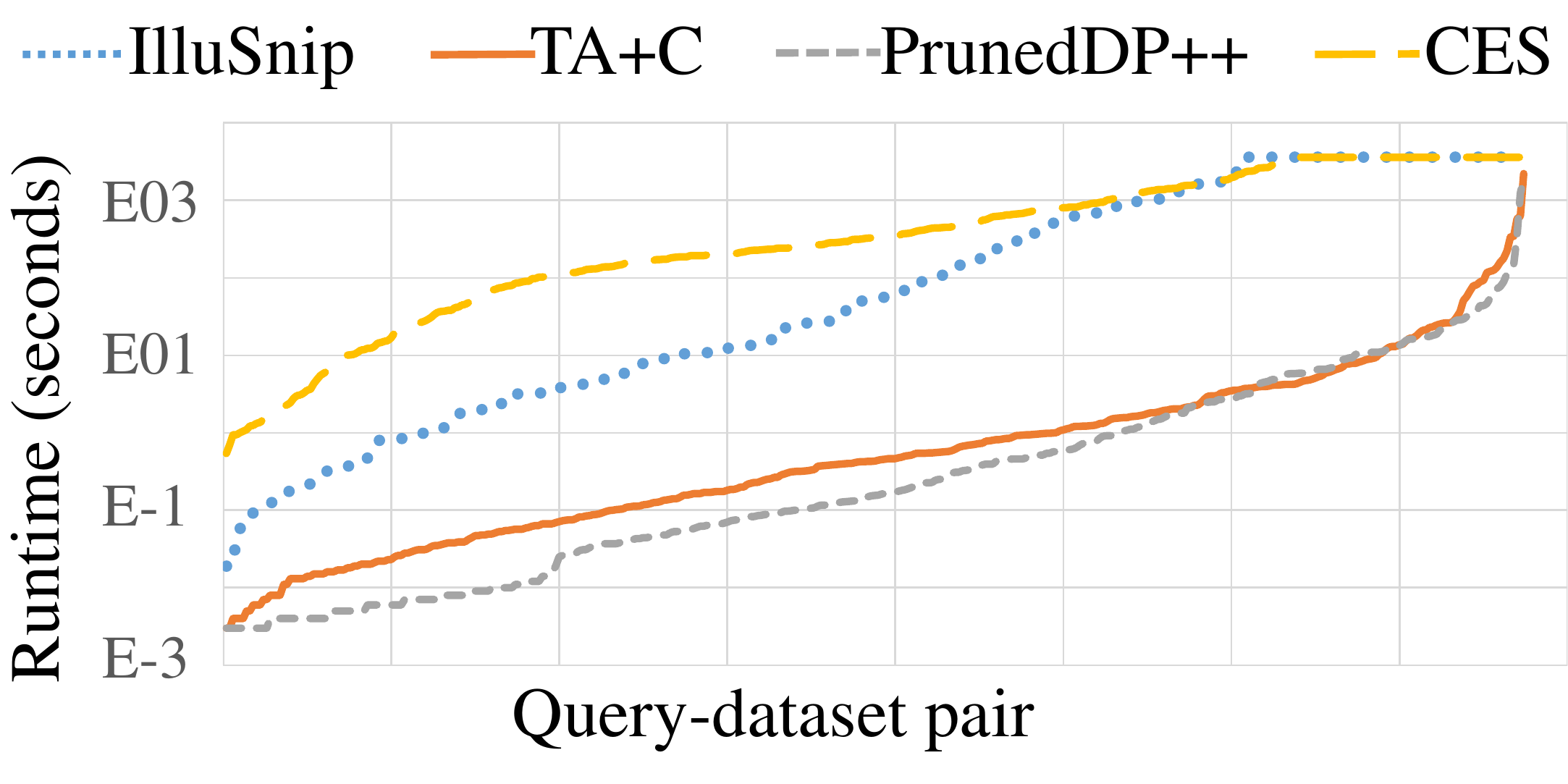}
        \caption{Runtime on each query-dataset pair, in ascending order.}
        \label{fig:runtime}
    \end{minipage}
\end{figure}

\begin{figure}[!t]
    \centering
    \includegraphics[width=\textwidth]{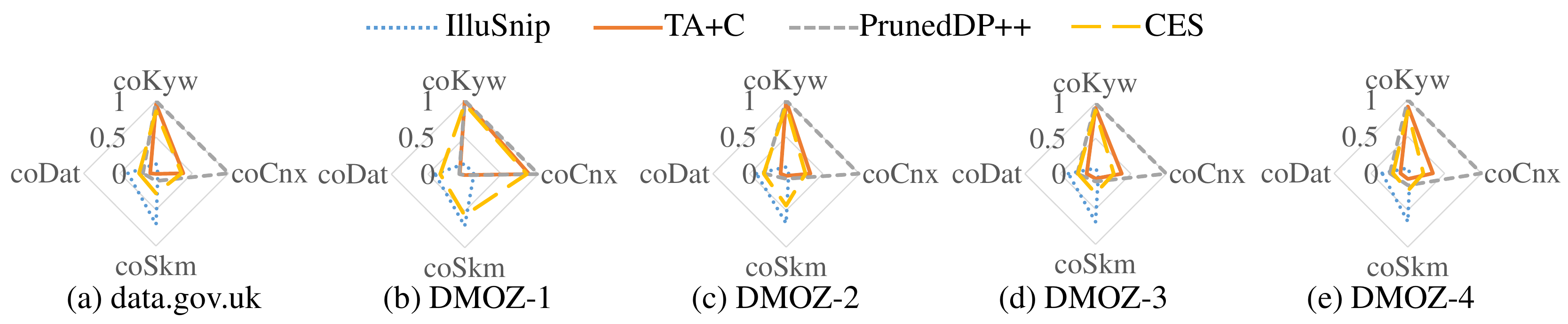}
    \caption{Average scores of evaluation metrics on each group of query-dataset pairs.}
    \label{fig:radar}
\end{figure}

Table~\ref{tab:avgsc} presents the average score of each evaluation metric achieved by each method on all the query-dataset pairs. Compared with Table~\ref{tab:teva}, a higher score was generally observed when a metric was conceptually considered in the components of a method. We concluded that the results of our empirical evaluation were basically consistent with our analysis in Section~\ref{sec:teva}. Figure~\ref{fig:radar} depicts the scores on each group of query-dataset pairs using radar charts.

\emph{IlluSnip}
achieved much higher scores of~$\cosch$ and~$\codat$ than other methods. It was not surprising because covering the schema and data of a dataset was central to the design of IlluSnip. However, there were still notable gaps between the achieved scores ($\cosch=0.6820$ and $\codat=0.3850$) and their upper bound (i.e.,~$1$), because IlluSnip was constrained to output a size-bounded connected subgraph. The coverage of such a subgraph was limited. On the other hand, all the other three methods were query-biased, whereas IlluSnip was not. Its very low scores of~$\cok=0.1000$ and~$\coc=0.0540$ suggested that the snippets generated by IlluSnip usually failed to cover queries.

\emph{TA$+$C}
was opposite in scores to IlluSnip. Coverage of dataset was not the focus of its design. The lowest scores of $\cosch=0.0425$ and $\codat=0.0915$  were observed on this method. By contrast, opting for connected subgraphs containing more query keywords, it achieved a fairly high score of $\cok=0.9590$.
%Still, as shown in Fig.~\ref{fig:radar}(b)--(e), when queries became longer, $\coc$~was falling because a radius-bounded connected subgraph was unlikely to cover many query keywords which could be distantly located in a dataset.
However, connections between query keywords were not captured well, because radius-bounded connected subgraph was incapable of covering long-distance connections. As shown in Fig.~\ref{fig:radar}, actually the overall score of $\coc=0.4703$ was even exaggerated by the case of DMOZ-1, where most queries comprised only one keyword and hence $\coc$~was trivially defined to be~$\cok$ according to Eq.~(\ref{eq:coc}). In other cases, $\coc$~was not high.

\emph{PrunedDP$++$}
could not find any GST to connect all the query keywords on $86$~query-dataset pairs, which had been excluded from our results. On the remaining pairs, not surprisingly, its coverage of query keywords ($\cok=1$) and their connections ($\coc=1$) was perfect. In a GST, query keywords were often connected via paths that passed through hub nodes in a dataset. Involving such high-degree nodes, a GST's coverage of data ($\codat=0.2133$) was considerably higher than that of TA$+$C ($\codat=0.0915$). However, similar to TA$+$C, a GST's coverage of data schema was limited ($\cosch=0.0898$).

\emph{CES}
appeared to be a more balanced method, as visualized in Fig.~\ref{fig:radar}. Towards generating a query-biased and diversified snippet, its coverage of query keywords ($\cok=0.9006$) was close to TA$+$C and PrunedDP$++$, and its coverage of dataset ($\cosch = 0.3668$ and $\codat = 0.2684$) was notably better. However, similar to TA$+$C, its coverage of connections between query keywords was not satisfying because selecting diversified triples usually led to a fragmented snippet. The overall score of $\coc=0.3926$ was exaggerated by the case of DMOZ-1.

\subsubsection{Runtime.}

We also evaluated the runtime of each method because fast generation of snippets is an expectation of search engine users. Figure~\ref{fig:runtime} depicts, on a logarithmic scale, the runtime of each method used for generating a snippet for each of the $387$~query-dataset pairs. Runtime was mainly related to the number of triples in a dataset.

PrunedDP$++$~was generally the fastest method, with a median runtime of $0.16$~second. It rarely reached timeout, and it completed computation in less than one second in 68\%~of the cases. TA$+$C was also reasonably fast, with a median runtime of $0.43$~second. These two methods showed promising performance for practical use. By contrast, IlluSnip and CES reached timeout in~$22\%$ and $18\%$~of the cases, respectively. They often spent tens or hundreds of seconds generating a snippet. Fortunately, IlluSnip was not query-biased, and hence could be used to generate snippets offline.
\section{User Study}\label{sec:us}

%Complementary to the above evaluation based on the proposed framework,
We recruited 20~students majoring in computer science to assess the quality of snippets generated by different methods. All the participants had the experience in working with RDF datasets. The results could be compared with the above evaluation results, to verify the effectiveness of our proposed evaluation metrics.

\subsubsection{Design.}

From fifty random candidates, each participant chose 5~datasets with interest according to their metadata.
% the 10 optional datasets were picked out randomly from the 30 least selected ones.
For each dataset, the participant had access to a list of classes and properties used in the dataset to help understanding.
% Then user was shown title, description and up to 100 labels extracted randomly from all classes and properties of the dataset selected.
The participant was required to describe some data needs that could be fulfilled by the dataset, and then repeatedly rephrase the needs as a keyword query until all of IlluSnip, TA$+$C, and PrunedDP$++$ could generate a non-empty snippet.
% They were required to write down 3 data requests, which are pieces of text describing the data they need in the dataset. Requests must be submitted before next step and users were not allowed to modify them after that. Then user was required to generate a query based on one of the 3 requests
% it doesn’t contain stop words; not all of its prefixes will lead to an empty snippet generated by PrunedDP++. The last condition means: if the snippet generated by prunedDP++ is empty, we will remove the last word of query and try again. This process is repeated until a non-empty snippet is generated (it meets the condition), or 0 word remains (it doesn’t meet the condition). User was required to modify the query if not all conditions meet.
For reasonable response time, CES was excluded from user study, and datasets containing more than one million triples were not used.
% We picked out datasets with 200 to 1000,000 triples, totally 176
Following~\cite{wsdm,ipm}, we visualized a snippet (which was an RDF graph) as a node-link diagram. The participant rated its usefulness in relevance judgment on a scale of 1--5, and commented its strengths and weaknesses.
% Besides, we requested every user to write a conclusive comment after reading all three snippets as a brief assessment. Each user was required to complete $5$ query-and-assess tasks independently.

%\begin{table}[!t]
%    \centering
%    \caption{Usefulness of Snippets (1--5) in Relevance Judgment}\label{tab:usr}
%    \begin{tabular}{|c|c|c|c|c|c|}
%        \hline
%        \multicolumn{3}{|c|}{Mean $\pm$ standard deviation} & rANOVA & LSD post-hoc \\
%        \cline{1-3}
%        IlluSnip & TA$+$C & PrunedDP$++$ & ($p$-value) & ($p<0.01$)\\
%        \hline
%        %Q1: It is relevant to the query. & $2.9$ & $2.9$ & $2.44$ & $1.99E-4$ &  IS, TA $>$ Pr \\
%        % It helped me to judge the usefulness of the dataset.
%        $3.10 \pm 1.28$ & $2.36 \pm 1.29$ & $1.92 \pm 1.19$ & $0.00264$ & IlluSnip $>$ TA$+$C, PrunedDP$++$ \\
%        \hline
%    \end{tabular}
%\end{table}

\subsubsection{Results.}

\begin{figure}[!t]
    \centering
    \begin{minipage}{.5\textwidth}
        \centering
        \captionsetup{type=table}
        \begin{tabular}{|c|c|c|}
            \hline
            \multicolumn{3}{|l|}{Mean $\pm$ standard deviation:}\\
            \hline
            IlluSnip & TA$+$C & PrunedDP$++$\\
            $3.10 \pm 1.28$ & $2.36 \pm 1.29$ & $1.92 \pm 1.19$\\
            \hline
            \hline
            \multicolumn{3}{|l|}{rANOVA ($p$-value): 0.00264}\\
            \hline
            \hline
            \multicolumn{3}{|l|}{LSD post-hoc ($p<0.01$):}\\
            \multicolumn{3}{|l|}{IlluSnip $>$ TA$+$C, PrunedDP$++$}\\
            \hline
        \end{tabular}
        \caption{Human-rated usefulness of snippets ($1$--$5$) in relevance judgment.}
        \label{tab:usr}
    \end{minipage}%
    \hfill
    \begin{minipage}{0.48\textwidth}
        \centering
        \includegraphics[width=\textwidth]{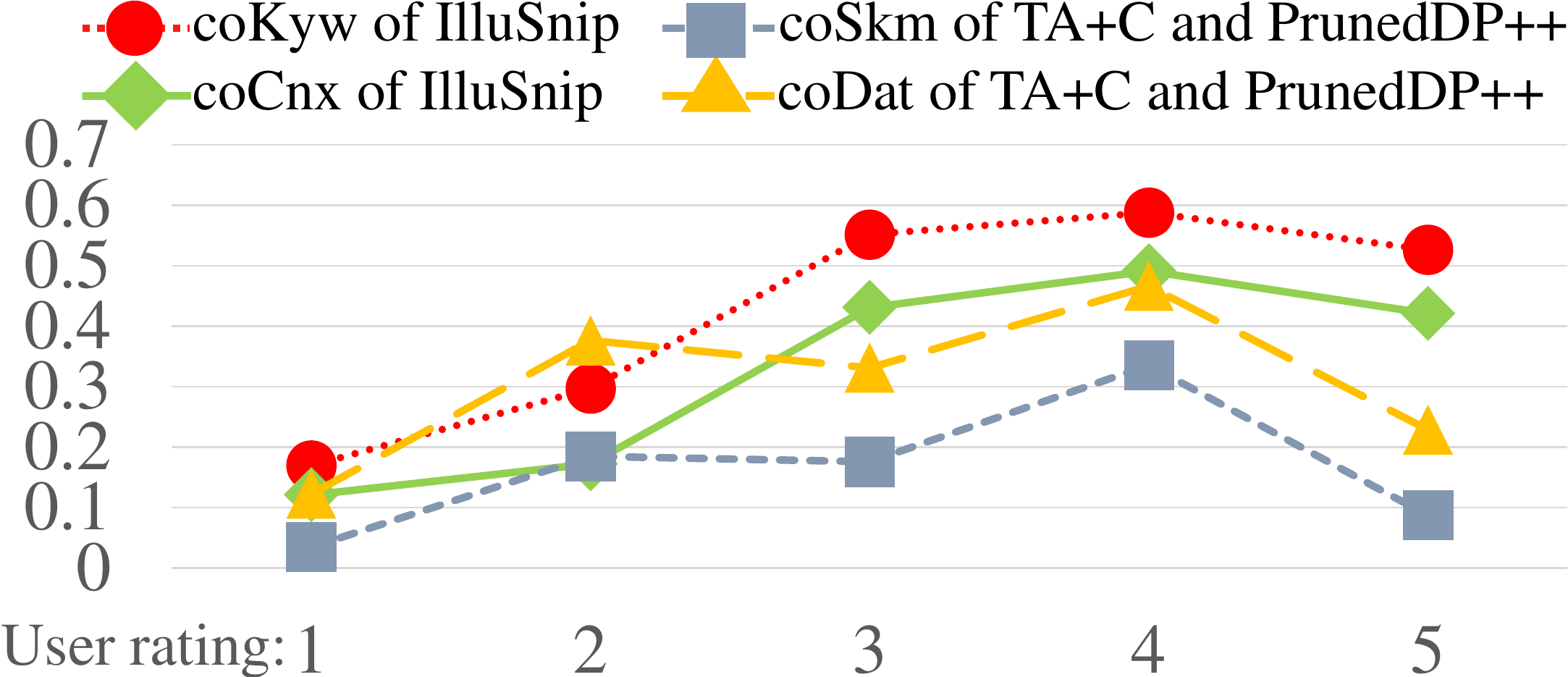}
        \caption{Correlation between evaluation metrics and user ratings.}
        \label{fig:mean}
    \end{minipage}
\end{figure}

Table~\ref{tab:usr} summarizes the responses from participants about snippets for a total of $20 \cdot 5 = 100$~datasets. IlluSnip received a higher mean rating than TA$+$C and PrunedDP$++$. Repeated measures ANOVA~(rANOVA) indicated that their differences were statistically significant ($p<0.01$). LSD post-hoc tests~($p<0.01$) suggested that IlluSnip was more helpful to users than TA$+$C and PrunedDP$++$, whereas the difference between TA$+$C and PrunedDP$++$ was not statistically significant.

Figure~\ref{fig:mean} shows the mean score of each evaluation metric, grouped by user ratings. For each evaluation metric we excluded the results of some methods when their scores were hardly distinguishable (all close to~1) because those methods had components that were conceptually similar to the metrics (cf.~Table~\ref{tab:teva}). The scores of all the four metrics generally increased as user ratings increased. The observed positive correlation demonstrated the effectiveness of our evaluation framework. Exceptions were the notable falls of~$\cosch$ and~$\codat$ at the end, where very few~($<10$) snippets were rated~5 so that the scores at this point might not be significant.
% which showed the potential to be used to support offline evaluation not involving human users, to supplement and even partially replace user study.

We analyzed participants' comments. For IlluSnip, $15$~participants~($75\%$) complimented the connectivity of its results which facilitated understanding, and $13$~participants~($65\%$) referred to the richness and diversity of the content, which accorded well with its high coverage of data schema. Not surprisingly, $16$~participants~($80\%$) criticized its weak relevance to the query. For TA$+$C, $15$~participants~($75\%$) appreciated its query relevance, but $19$~participants~($95\%$) complained that its results sometimes contained many isolated nodes. It happened frequently when a query contained only one keyword. Although these nodes covered the query keyword, they were not associated with any further description, which dissatisfied $12$~participants~($60\%$). For PrunedDP$++$, similar feedback was received from $17$~participants~($85\%$) for some cases, but in other cases, $15$~participants~($75\%$) commented its high coverage of query keywords and the paths between them, which facilitated the comprehension of their connections. Besides, $8$~participants~($40\%$) favored the conciseness of its results.

Participants were invited to a post-experiment interview. Whereas they confirmed the usefulness of snippets, they generally believed that snippet could not replace but complement abstractive summary with statistics. Some participants suggested implementing interactive (e.g.,~hierarchical and zoomable) snippets for user exploration, which could be a future direction of research.

\subsubsection{Discussion.}

Participants' ratings, comments, and the results of our proposed evaluation metrics were generally consistent with each other. The results justified the appropriateness of our framework to evaluating snippets in dataset search.

From the participants' comments, we concluded that a good dataset snippet should, on the one hand, cover query keywords and their connections to make sense of the underlying query intent. TA$+$C and PrunedDP$++$ were focused on this aspect. On the other hand, it should provide rich and diverse description about matched resources and triples to make sense of the dataset content. This was overlooked by TA$+$C and PrunedDP$++$, and it suggested a difference between snippet generation and keyword search. A trade-off between informativeness and compactness should be considered. IlluSnip showed promising results along this way. However, none of the three methods fulfilled these requirements completely, and hence their usefulness scores were far from perfection.
% Secondly, a satisfying snippet should possess appropriate connectivity. Most participants of our experiment preferred connected and compact snippets than those having loose structures which occurred frequently in results of TA$+$C, indicating the connectivity of snippets could benefit users' comprehension to relationships between keywords and the dataset. But on the other hand, too many connections between same keywords may bring in redundancy information, which should be avoided in practice.
\section{Conclusion}\label{sec:cfd}

To promote research on the emerging problem of snippet generation for dataset search, we have proposed an evaluation framework for assessing the quality of dataset snippets. With our metrics, methods proposed in the future can be evaluated more easily. Our framework relies on neither ground-truth snippets which are difficult to create, nor human efforts in user study which are inefficient and expensive. Evaluation can be automated offline. This in turn will be beneficial to the rapid development and deployment of snippets for dataset search engines.

Our evaluation results reveal the shortcomings of state-of-the-art methods adapted from related fields, which are also verified by a user study. None of the evaluated methods address all the considered aspects. It inspires us to put forward new methods for generating dataset snippets with more comprehensive features. Efficiency and scalability of methods are also important factors. Storage will also be a concern because a dataset search engine may have to store and index each dataset for snippet generation.
% The quality of snippets generated for dataset search depends on many separated aspects. There might be other standards to assess the usefulness of snippets apart from our proposed metrics. For example, when we are dealing with two or more large similar datasets, an approach is needed to effectively distinguish them in content. In other word, a metric for assessing the distinctive characteristics of a dataset can be added to our framework. We are still working on improving our measurements and adding more metrics to our evaluation system.

Our work has the following limitations. First, our evaluation framework may not be comprehensive. It can partially assess the quality of a dataset snippet, but still is not ready to completely replace user study. There may be other useful metrics, such as distinctiveness, readability, and coherence, which we will study in future work. Second, our evaluation metrics are implemented specifically for RDF datasets. To extend the range of application of our framework, more generalized implementation for other data formats needs to be explored.
% Third, we define a snippet as an extract from a dataset. By contrast, existing dataset search engines show some metadata about a dataset as a snippet. These two types of snippets should be complementary. In practice, the keywords in a search query may refer to both of them. However, metadata is not considered in our work. Our evaluation framework needs extension when a more general definition of snippet is used.

\subsection*{Acknowledgements}
This work was supported in part by the National Key R\&D Program of China under Grant 2018YFB1005100, in part by the NSFC under Grant 61572247, and in part by the SIRIUS Centre, Norwegian Research Council project number 237898. Cheng was funded by the Six Talent Peaks Program of Jiangsu Province under Grant RJFW-011.

\end{document}